# Octonions, $G_2$ Symmetry, Generalized Self-Duality and Supersymmetries in Dimensions $D \leq 8$ [1]

Hitoshi NISHINO[2] and Subhash RAJPOOT[3]

*Department of Physics & Astronomy*
*California State University*
*1250 Bellflower Blvd.*
*Long Beach, CA 90840*

## Abstract

We establish $N = (1/8, 1)$ supersymmetric Yang-Mills vector multiplet with generalized self-duality in Euclidian eight-dimensions with the original full $SO(8)$ Lorentz covariance reduced to $SO(7)$. The key ingredient is the usage of octonion structure constants made compatible with $SO(7)$ covariance and chirality in 8D. By a simple dimensional reduction together with extra constraints, we derive $N = 1/8 + 7/8$ supersymmetric self-dual vector multiplet in 7D with the full $SO(7)$ Lorentz covariance reduced to $G_2$. We find that extra constraints needed on fields and supersymmetry parameter are not obtained from a simple dimensional reduction from 8D. We conjecture that other self-dual supersymmetric theories in lower dimensions $D = 6$ and $4$ with respective reduced global Lorentz covariances such as $SU(3) \subset SO(6)$ and $SU(2) \subset SO(4)$ can be obtained in a similar fashion.

---

[1]Work supported in part by NSF Grant # 0308246
[2]E-Mail: hnishino@csulb.edu
[3]E-Mail: rajpoot@csulb.edu

## 1. Introduction

The recent studies of M-theory [1] lead to an important conclusion that in order for realistic four-dimensional (4D) theory with chiral fermions to emerge out of 11D supergravity, the extra 7D manifold should have the reduced holonomy $G_2$ instead of the maximal $SO(7)$. This is analogous to the case of superstring compactification from 10D into 4D with 6D Calabi-Yau manifold for with the reduced $SU(3)$ holonomy instead of the maximal $SO(6) \approx SU(4)$.

For manifolds of homogeneous spaces with reduced holonomy, such as 8D with $SO(7) \approx Spin(7)$ holonomy, or $G_2$, $SU(3)$, $SU(2)$ holonomies respectively in dimensions $D = 8, 7, 6$ and $4$, it is known that certain self-duality conditions for spin connection are playing important roles [2][3][4], such as leading to the desired holonomies of $Spin(7)$, $G_2$, $SU(3)$ and $SU(2)$, respectively.

Motivated by these developments, the importance of self-duality in dimensions $5 \leq D \leq 8$ has been more emphasized nowadays [5][6][3]. Symbolically, a self-duality condition in dimensions $5 \leq D \leq 8$ looks like $F_{\mu\nu} = (1/2)\phi_{\mu\nu}{}^{\rho\sigma} F_{\rho\sigma}$ where the usual $\epsilon$-tensor $\epsilon_{\mu\nu}{}^{\rho\sigma}$ in 4D is generalized into a 4-th rank antisymmetric tensor $\phi_{\mu\nu}{}^{\rho\sigma}$ which is constant under holonomies of certain manifolds, such as $SO(8)/SO(7) \approx Spin(8)/Spin(7)$ or $SO(7)/G_2 \approx Spin(7)/G_2$. Since this 'self-duality' is not between the usual Hodge-dual tensors, but only between the second-rank tensors in these dimensions, it is sometimes called generalized self-duality [3][5].

Self-dual supersymmetric vector multiplet in 4D [7] has been formulated with simple as well as extended supersymmetries [8][9]. The main ingredient that bridges self-duality and supersymmetry in 4D is the chirality of fermions leading to the self-duality of the field strength under supersymmetry. However, in the case of higher-dimensional self-dualities in $D = 8, 7, 6, 4$, we need something more than the usual chirality, as will be explained later. We show in this paper that certain constraints on component fields and the parameter of supersymmetry play such a role of supersymmetrizing generalized self-dualities in dimensions $D = 8$ and $7$ with covariances[4] under $Spin(7)$ and $G_2$.

At the bosonic field level, 'generalized' self-dualities in higher dimensions have been considered for some time [3][5]. There have been some important works in this direction, such as those by L. Baulieu *et al.* [10], and/or by Acharya *et al.* [11] with non-self-dual supersymmetry. Or exact solutions with Euclidean manifolds with reduced holonomies are obtained based on D-brane solutions or other compactifications [12]. However, to our knowledge, there has been no explicit formulation for self-dual supersymmetric Yang-Mills theories themselves in 8D or 7D with reduced symmetries $Spin(7)$ or $G_2$. The most important purpose of this paper is to fill this gap, *i.e.,* presenting the explicit component formulation of supersymmetric Yang-Mills theories in 8D or 7D with generalized self-duality, inspecting the mutual consistency of all the needed conditions with enough details, in particular, associated with the subtlety mentioned above.

---
[4]Here we use the term 'covariances' instead of holonomy, because these symmetries are global in this paper.

In the next section, we give the formulation of self-dual supersymmetric Yang-Mills theory with the reduced Lorentz covariance $Spin(7)$ in 8D. This is done by imposing certain supersymmetric self-duality conditions on supersymmetric Yang-Mills theory in 8D. In section 3, we first use the simple dimensional reduction to get the supersymmetric Yang-Mills theory in 7D, before imposing any self-duality. Adopting a prescription similar to 8D, we next give self-dual supersymmetric Yang-Mills theory in 7D with the reduced Lorentz covariance $G_2$. After Concluding Remarks, we give Appendix for a superspace reformulation of self-dual supersymmetric Yang-Mills theory in 8D, that might be of great help for understanding the consistency of our total system, in particular with the commutator algebra and consistency among constraints, together with the reduced Lorentz covariance $Spin(7)$ which is nontrivial and unconventional in supersymmetry.

## 2. Self-Dual $N = (1/8, 1)$ Supersymmetric Vector Multiplet in 8D

We start with a Yang-Mills vector multiplet in Euclidean 8D, *before* imposing generalized self-duality, as the foundation for self-dual case. Before imposing such a condition, the system has non-chiral $N = (1, 1)$ supersymmetry.[5] The spinor structures in such Euclidean 8D space can be easily understood by the general formulations in [13].

The field content of $N = (1, 1)$ Yang-Mills vector multiplet is $(A_\mu{}^I, \lambda_\alpha{}^I, B^I, C^I)$ with 8+8 degrees of freedom for each value of adjoint indices $I, J, \cdots = 1, 2, \cdots, \dim G$ for a gauged non-Abelian group $G$. The $\lambda$ is a 16-component Majorana spinor in Euclidean 8D [13], while the $B$ and $C$ fields are respectively real scalar and pseudo-scalars. We use $\mu, \nu, \cdots = 1, 2, \cdots, 8$ for the 8D space indices, and $\alpha, \beta, \cdots = 1, 2, \cdots, 16$ for spinorial indices for Majorana spinors in 8D, while our metric is $(\eta_{\mu\nu}) = \text{diag.}(+, +, \cdots, +)$ and $\{\gamma_\mu, \gamma_\nu\} = +2\eta_{\mu\nu}$, $\epsilon^{12\cdots 8} = +1$, $\gamma_9 \equiv \gamma_1 \gamma_2 \cdots \gamma_8$ [13].

Before fixing our lagrangian, we give some preliminaries for readers who are less familiar with spinors in 8D. On the basis of the known classification of possible supersymmetry algebras in various dimensions [13], the readers should note the following facts: The most crucial feature in 8D is that the charge-conjugation matrix is symmetric [13], which can be chosen to be the Kronecker's delta: $C_{\alpha\beta} = \delta_{\alpha\beta}$. Accordingly, we can choose the symmetries of the gamma matrices in 8D as in the following Table:

---

[5]The notation $N = (1, 1)$ means $N = 1$ left-handed ($\mathbf{8}_s$) and $N = 1$ right-handed ($\mathbf{8}_c$) supersymmetries. This is distinguished from the chiral supersymmetries $N = (1/8, 1)$ for self-dual supersymmetry described later. See also the next paragraph after eq. (2.24).

| $\gamma$-matrices | $I$ | $\gamma^\mu$ | $\gamma^{[2]}$ | $\gamma^{[3]}$ | $\gamma^{[4]}$ | $\gamma^{[5]}$ | $\gamma^{[6]}$ | $\gamma^{[7]}$ | $\gamma^{[8]}$ | Sum Total |
|---|---|---|---|---|---|---|---|---|---|---|
| # of Components | 1 | 8 | 28 | 56 | 70 | 56 | 28 | 8 | 1 | 256 |
| Symmetric | 1 | 8 | - | - | 70 | 56 | - | - | 1 | 136 |
| Anti-Symmetric | - | - | 28 | 56 | - | - | 28 | 8 | - | 120 |

Table 1: Numbers of Degrees of Freedom of $\gamma$-Matrices in Euclidean 8D

In Table 1, the numbers in the square brackets $\gamma^{[n]}$ for the $\gamma$-matrices are for $n$-th rank totally antisymmetric tensorial indices, *e.g.*, $\gamma^{[3]} \approx \gamma^{\mu\nu\rho}$. Now we see that we can choose both the identity matrix and $\gamma^\mu$ to be symmetric in 8D. Of course, we can have an alternative option, such as $\gamma^{[n]}$ and $\gamma^{[8-n]}$ are completely flipped around, so that $I$, $\gamma^{[3]}$, $\gamma^{[4]}$, $\gamma^{[7]}$, $\gamma^{[8]}$, are now symmetric with total 136 components, while the remaining 120 components are anti-symmetric [13]. However, in such a case, $\gamma^\mu$ becomes anti-symmetric, so that we have the spinorial inner product $(\overline\psi\gamma^\mu\chi) = +(\overline\chi\gamma^\mu\psi)$, which is not desirable for supersymmetry commutator algebra, because we need $(\overline\epsilon_1\gamma^\mu\epsilon_2) = -(\overline\epsilon_2\gamma^\mu\epsilon_1)$ for two supersymmetry transformations $\delta_Q(\epsilon_1)$ and $\delta_Q(\epsilon_2)$. For this reason, we follow the first option as in Table 1 so that $\gamma^\mu$ is symmetric, and therefore $(\overline\psi\gamma^\mu\chi) = -(\overline\chi\gamma^\mu\psi)$, as desired. As a matter of fact, we can also reformulate the whole system in superspace, as will be performed in Appendix as a good cross examination.[6]

Upon this assignment of $\gamma$-matrix symmetry, general spinorial inner products have the properties $(\overline\psi\gamma^{\mu_1\cdots\mu_n}\chi) = -(-1)^{n(n-1)/2}(\overline\chi\gamma^{\mu_1\cdots\mu_n}\psi)$ and $(\overline\psi\gamma^{\mu_1\cdots\mu_n}\chi)^\dagger = -(\overline\psi\gamma^{\mu_1\cdots\mu_n}\chi)$ [13]. This is due to the properties such as $(\gamma_\mu)^\dagger = +\gamma_\mu$, $\overline\psi \equiv \psi^\dagger$, $\psi^* = \psi$, because both $A$ and $B$-matrices in [13] can be an unit matrix. As has been mentioned, since $C_{\alpha\beta} = \delta_{\alpha\beta}$, spinorial indices need no distinctions for raising/lowering, *e.g.*, $(\overline\psi\gamma^\mu\chi) \equiv \psi_\alpha(\gamma^\mu)_{\alpha\beta}\chi_\beta$.

We are now ready to present our invariant action $I_{8\mathrm D} \equiv \int d^8x\, \mathcal{L}_{8\mathrm D}$ with the lagrangian[7]

$$\mathcal{L}_{8D} = -\tfrac{1}{4}(F_{\mu\nu}{}^I)^2 - i(\overline\lambda^I\gamma^\mu D_\mu\lambda^I) + \tfrac{1}{2}(\partial_\mu B^I)^2 - \tfrac{1}{2}(\partial_\mu C^I)^2$$
$$- if^{IJK}B^I(\overline\lambda^J\lambda^K) + if^{IJK}C^I(\overline\lambda^J\gamma_9\lambda^K) + \tfrac{1}{2}(f^{IJK}B^JC^K)^2 \quad, \qquad (2.1)$$

invariant under $N = (1,1)$ supersymmetry

$$\delta_Q A_\mu{}^I = +i(\overline\epsilon\gamma_\mu\lambda^I) \quad, \quad \delta_Q B^I = +i(\overline\epsilon\lambda^I) \quad, \quad \delta_Q C^I = +i(\overline\epsilon\gamma_9\lambda) \quad, \qquad (2.2\mathrm a)$$
$$\delta_Q \lambda^I = -\tfrac{1}{4}(\gamma^{\mu\nu}\epsilon)F_{\mu\nu}{}^I + \tfrac{1}{2}(\gamma^\mu\epsilon)D_\mu B^I + \tfrac{1}{2}(\gamma_9\gamma^\mu\epsilon)D_\mu C^I - \tfrac{1}{2}f^{IJK}(\gamma_9\epsilon)B^JC^K \quad. \qquad (2.2\mathrm b)$$

Here $f^{IJK}$ are the structure constants of the gauge group $G$.

---
[6] Note also that we can formulate self-dual supergravity in superspace based on the same notation, as in [14]

[7] This lagrangian itself is not a new result. In fact, it coincides with a lagrangian given in the first paper in [11] which was obtained in a BRST formulation.

The closure of our supersymmetry is easily confirmed on each field, as

$$[\delta_Q(\epsilon_1), \delta_Q(\epsilon_2)] = \delta_P(\xi^\mu) \ , \quad \xi^\mu \equiv +i(\overline{\epsilon}_2 \gamma^\mu \epsilon_1) \ , \tag{2.3}$$

where $\delta_P$ is the translation operator as usual in supersymmetric theories.

In the superinvariance confirmation of our action, we encounter the cancellation of $\lambda^3$-terms, that need the Fierz identity $[\,(\overline{\psi}_1 \gamma_\mu \psi_2)(\overline{\psi}_3 \gamma^\mu \psi_4) - (\overline{\psi}_1 \psi_2)(\overline{\psi}_3 \psi_4) + (\overline{\psi}_1 \gamma_9 \psi_2)(\overline{\psi}_3 \gamma_9 \psi_4)\,]$ $+ (2 \text{ perms.}) \equiv 0$, where (2 perms.) is needed for the total antisymmetrization under $2 \leftrightarrow 3 \leftrightarrow 4$.

The field equations of our system are now

$$\slashed{D}\lambda^I + f^{IJK}\lambda^J B^K - f^{IJK}\gamma_9 \lambda^J C^K \doteq 0 \ , \tag{2.4a}$$

$$D_\nu F_\mu{}^{\nu I} - if^{IJK}(\overline{\lambda}^J \gamma_\mu \lambda^K) - f^{IJK} B^J D_\mu B^K + f^{IJK} C^J D_\mu C^K \doteq 0 \ , \tag{2.4b}$$

$$D_\mu^2 B^I + if^{IJK}(\overline{\lambda}^J \lambda^K) - f^{IJK} f^{KLM} B^L C^J C^M \doteq 0 \ , \tag{2.4c}$$

$$D_\mu^2 C^I + if^{IJK}(\overline{\lambda}^J \gamma_9 \lambda^K) - f^{IJK} f^{KLM} B^J B^L C^M \doteq 0 \ , \tag{2.4d}$$

where $\doteq$ symbolizes a field equation distinct from identities.

We are now ready to consider generalized self-duality [3][5] consistent with supersymmetry in 8D. The most important ingredients are the definitions of the 4-th rank antisymmetric $SO(7)$-invariant constant $f_{\mu\nu\rho\sigma}$ defined in terms of octonion structure constants $\psi_{\mu\nu\rho}$ [6]:

$$\psi_{123} = \psi_{516} = \psi_{624} = \psi_{435} = \psi_{471} = \psi_{673} = \psi_{572} = +1 \ ,$$

$$\phi_{4567} = \phi_{2374} = \phi_{1357} = \phi_{1276} = \phi_{2356} = \phi_{1245} = \phi_{1346} = +1 \ ,$$

$$\phi_{\mu\nu\rho\sigma} \equiv +(1/3!)\epsilon^{\mu\nu\rho\sigma\tau\omega\psi}\psi_{\tau\omega\psi} \ . \tag{2.5}$$

Thus $\phi_{\mu\nu\rho\sigma}$ and $\psi_{\mu\nu\rho}$ are dual to each other. Now $f_{\mu\rho\sigma\tau}$ are defined by

$$f_{\mu\nu\rho 8} \equiv \psi_{\mu\nu\rho} \ , \quad f_{\mu\nu\rho\sigma} \equiv \phi_{\mu\nu\rho\sigma} \quad (\text{ for } \mu, \nu, \rho = 1, 2, \cdots, 7) \ , \tag{2.6}$$

so that explicitly, we have

$$f_{1238} = f_{5168} = f_{6248} = f_{4358} = f_{4718} = f_{6738} = f_{5728} = +1 \ ,$$

$$f_{4567} = f_{2374} = f_{1357} = f_{1276} = f_{2356} = f_{1245} = f_{1346} = +1 \ . \tag{2.7}$$

Accordingly, they also satisfy the self-duality in 8D:

$$f^{\mu\nu\rho\sigma} = +\tfrac{1}{4!}\epsilon^{\mu\nu\rho\sigma\tau\lambda\omega\psi} f_{\tau\lambda\omega\psi} \ . \tag{2.8}$$

Even though $f_{\mu\nu\rho\sigma}$ is not covariant under the full $SO(8) \approx Spin(8)$, it is covariant under its subgroup $SO(7) \approx Spin(7)$. Accordingly, the system has now the covariance under $SO(7) \approx Spin(7)$ reduced from the full $SO(8) \approx Spin(8)$. In other words, the usual concept of 'Lorentz covariance' has also a reduced meaning.

Several useful relationships associated with the $f$'s are such as[8] [6]

$$f_{\mu\nu\rho\sigma} f^{\sigma\tau\lambda\omega} = -6\delta_{[\mu}{}^\tau \delta_\nu{}^\lambda \delta_{\rho]}{}^\omega + 9 f_{[\mu\nu}{}^{[\tau\lambda} \delta_{\rho]}{}^{\omega]} ,$$

$$f_{\mu\nu\rho\sigma} f^{\rho\sigma\tau\lambda} = +12\delta_{[\mu}{}^\tau \delta_{\nu]}{}^\lambda - 4 f_{\mu\nu}{}^{\tau\lambda} ,$$

$$f_{\mu[3]} f^{\nu[3]} = +42 \delta_\mu{}^\nu , \quad (f_{[4]})^2 = +336 . \tag{2.9}$$

The symbol $[n]$ denotes totally antisymmetric $n$ indices. The chirality projectors in 8D

$$P \equiv \frac{I+\gamma_9}{2} , \quad N \equiv \frac{I-\gamma_9}{2} , \tag{2.10}$$

and the projectors [6][3][5]

$$P_{\mu\nu}{}^{\rho\sigma} \equiv \tfrac{3}{4}\left(\delta_{[\mu}{}^\rho \delta_{\nu]}{}^\sigma + \tfrac{1}{6} f_{\mu\nu}{}^{\rho\sigma}\right) , \quad N_{\mu\nu}{}^{\rho\sigma} \equiv \tfrac{1}{4}\left(\delta_{[\mu}{}^\rho \delta_{\nu]}{}^\sigma - \tfrac{1}{2} f_{\mu\nu}{}^{\rho\sigma}\right) , \tag{2.11}$$

with the $\gamma$-matrix combination

$$f \equiv \tfrac{1}{4!} f^{\mu\nu\rho\sigma} \gamma_{\mu\nu\rho\sigma} , \tag{2.12}$$

we get useful relationships, such as

$$f = f\gamma_9 = \gamma_9 f , \quad f\gamma_\mu = -\tfrac{1}{3} f_\mu{}^{[3]} P \gamma_{[3]} , \quad \gamma_\mu f = +\tfrac{1}{3} f_\mu{}^{[3]} N \gamma_9 \gamma_{[3]} ,$$

$$P_{\mu\nu}{}^{\rho\sigma} P_{\rho\sigma}{}^{\tau\lambda} = P_{\mu\nu}{}^{\tau\lambda} , \quad N_{\mu\nu}{}^{\rho\sigma} N_{\rho\sigma}{}^{\tau\lambda} = N_{\mu\nu}{}^{\tau\lambda} , \quad P_{\mu\nu}{}^{\rho\sigma} N_{\rho\sigma}{}^{\tau\lambda} = N_{\mu\nu}{}^{\rho\sigma} P_{\rho\sigma}{}^{\tau\lambda} = 0 ,$$

$$\mathcal{P} \equiv \tfrac{1}{8}(P + \tfrac{1}{2} f) , \quad \mathcal{Q} \equiv \tfrac{7}{8}(P - \tfrac{1}{14} f) , \quad \mathcal{P}^2 = \mathcal{P} , \quad \mathcal{Q}^2 = \mathcal{Q} , \quad \mathcal{P}\mathcal{Q} = \mathcal{Q}\mathcal{P} = 0 . \tag{2.13}$$

Armed with these relationships, we are ready to formulate supersymmetric generalized self-duality in 8D. Following the pattern with the self-duality in 4D [8][9], we expect such generalized self-duality is closely associated with chirality in 8D. The simplest choice is just to impose the conditions $P\lambda = 0$ or $N\lambda = 0$. Now, as the flipping property in 8D shows, e.g., $(\overline{\psi}\chi) = -(\overline{\chi}\psi)$, $(\overline{\psi}\gamma_\mu\chi) = -(\overline{\chi}\gamma_\mu\psi)$, $(\overline{\psi}\gamma_9\chi) = -(\overline{\chi}\gamma_9\psi)$, the product $(\overline{\epsilon}\gamma_\mu\lambda)$ in (2.2a) makes sense, only if $\epsilon$ and $\lambda$ have *opposite* chiralities. However, the problem is (2.2b), showing that both $\epsilon$ and $\lambda$ should have the *same* chirality. In other words, imposing only the chirality condition is not enough for supersymmetric generalized self-duality. On the other hand, we also anticipate that chirality is crucial from experience in 4D [8][9].

What we need is not only the chirality condition[9]

$$\lambda_+{}^I \equiv P\lambda^I \equiv \frac{I+\gamma_9}{2} \lambda^I \stackrel{*}{=} 0 , \tag{2.14}$$

---

[8]Even though we are in Euclidean 8D, it is sometimes convenient to use both super and subscripts in equations like (2.9) with multiple anti-symmetrizations.

[9]We use the symbol $\stackrel{*}{=}$ for conditions related to generalized self-duality

but also certain constraint on the parameter of supersymmetry $\epsilon_+$. In fact, it has been known [3] that the parameter $\epsilon$ is subject to Killing spinor condition on our particular $SO(7)$ as a subgroup of the total $SO(8)$ Lorentz covariance $P_{\mu\nu}{}^{\rho\sigma}\gamma_{\rho\sigma}P\epsilon \equiv \gamma^{\mu\nu}_{(+)}\epsilon_+ \stackrel{*}{=} 0$, which is the necessary and sufficient condition of $\mathcal{Q}\epsilon_+ \stackrel{*}{=} 0$:

$$\mathcal{Q}\,\epsilon_+ = \tfrac{7}{8}(P - \tfrac{1}{14}f)\,\epsilon_+ \stackrel{*}{=} 0 \iff \gamma^{(+)}_{\mu\nu}\epsilon_+ \stackrel{*}{=} 0 \ , \tag{2.15}$$

Once (2.14) and (2.15) are imposed, we can take the variation of the former under supersymmetry (2.2) to get all other bosonic field equations:

$$0 \stackrel{*}{=} \delta_Q \lambda^I_+ \stackrel{*}{=} -\tfrac{1}{4}\gamma^{\mu\nu}_{(-)}\,\epsilon_+ N_{\mu\nu}{}^{\rho\sigma}F_{\rho\sigma}{}^I + \tfrac{1}{2}\gamma^\mu \epsilon_- D_\mu(B^I + C^I) - \tfrac{1}{2}\epsilon_+ f^{IJK}B^J C^K \ . \tag{2.16}$$

Demanding each term of different $\gamma$-matrix structures to vanish yields

$$N_{\mu\nu}{}^{\rho\sigma}F_{\rho\sigma}{}^I \stackrel{*}{=} 0 \iff F_{\mu\nu}{}^I \stackrel{*}{=} +\tfrac{1}{2}f_{\mu\nu}{}^{\rho\sigma}F_{\rho\sigma} \ , \tag{2.17a}$$

$$D_\mu(B^I + C^I) \stackrel{*}{=} 0 \iff B^I \stackrel{*}{=} -C^I \ . \tag{2.17b}$$

Eq. (2.17b) is due to gauge covariance under the gauge group $G$.

The mutual consistency among the conditions (2.17), (2.14), (2.15) with supersymmetry (2.2) is confirmed as follows: First of all, the variation of (2.17b) under (2.18) is easily shown to vanish under (2.14). Next, the supersymmetric variation of (2.17a) is

$$
\begin{aligned}
0 \stackrel{?}{=} \delta_Q(N_{\mu\nu}{}^{\rho\sigma}F_{\rho\sigma}{}^I) &\stackrel{*}{=} -2i N_{\mu\nu}{}^{\rho\sigma}(\overline{\epsilon}_+\gamma_\rho D_\sigma \lambda_-{}^I) \stackrel{*}{=} -\tfrac{i}{7}N_{\mu\nu}{}^{\rho\sigma}(\overline{\epsilon}_+ f\gamma_\rho D_\sigma \lambda_-{}^I) \\
&\stackrel{*}{=} -\tfrac{i}{7}N_{\mu\nu}{}^{\rho\sigma}(\overline{\epsilon}_+[f,\gamma_\rho]D_\sigma \lambda_-{}^I) - \tfrac{i}{7}N_{\mu\nu}{}^{\rho\sigma}\left(\overline{\epsilon}_+\gamma_\rho D_\sigma(f\lambda_-{}^I)\right) \\
&\stackrel{*}{=} +\tfrac{i}{84}f_{[\mu|}{}^{[3]}(\overline{\epsilon}_+\gamma_{[3]}D_{|\nu]}\lambda_-{}^I) - \tfrac{i}{168}(+6\delta_\mu{}^\tau \delta_\nu{}^\lambda \delta_\sigma{}^\omega - 9 f_{[\mu\nu}{}^{\tau\lambda}\delta_{\sigma]}{}^\omega)(\overline{\epsilon}_+\gamma_{\tau\lambda\omega}D^\sigma\lambda_-{}^I) \\
&\stackrel{*}{=} +\tfrac{i}{84}f_{[\mu|}{}^{[3]}(\overline{\epsilon}_+\gamma_{[3]}D_{|\nu]}\lambda_-{}^I) + \tfrac{i}{14}(\overline{\epsilon}_+\gamma_{[\mu}D_{\nu]}\lambda_-{}^I) - \tfrac{i}{28}(\overline{\epsilon}_+ f_{\mu\nu}{}^{\rho\sigma}\gamma_\rho D_\sigma\lambda_-{}^I) \\
&\quad -\tfrac{i}{28}f_{[\mu|}{}^{[2]\rho}(\overline{\epsilon}_+\gamma_{|\nu][2]}D_\rho\lambda_-{}^I) \\
&\stackrel{*}{=} +\tfrac{2i}{7}N_{\mu\nu}{}^{\rho\sigma}(\overline{\epsilon}_+\gamma_\rho D_\sigma\lambda_-{}^I) - \tfrac{i}{21}f_{[\mu}{}^{\rho\sigma\tau}A_{\nu]\rho\sigma\tau} \ ,
\end{aligned}
\tag{2.18}
$$

where $A_{\mu\nu\rho\sigma} \equiv (\overline{\epsilon}_+\gamma_{[\mu\nu\rho}D_{\sigma]}\lambda_-{}^I)$, and we used frequently the relationships (2.11) - (2.13) leading to $f\lambda_-{}^I = fN\lambda_-{}^I \equiv 0$, constraints (2.14) and (2.16), with the $\lambda$-field equation (2.4a) with (2.17b):

$$\gamma^\mu D_\mu \lambda_-{}^I \stackrel{.}{=} 0 \ . \tag{2.19}$$

Since the first term in the last side of (2.18) is proportional to its second side with a non-unity coefficient, all we have to show is that the vanishing of the last term in (2.18) on-shell. First, we show that $A_{\mu\nu\rho\sigma}$ is on-shell anti-self-dual: $A_{[4]} \stackrel{.}{=} -(1/4!)\,\epsilon_{[4]}{}^{[4]'}A_{[4]'}$ which can be easily done, once we can show that

$$\gamma_{[\mu\nu\rho}D_{\sigma]}\lambda_-{}^I \stackrel{.}{=} -\tfrac{1}{24}\epsilon_{\mu\nu\rho\sigma}{}^{\tau\lambda\omega\psi}\gamma_{\tau\lambda\omega}D_\psi\lambda_-{}^I \ , \tag{2.20}$$

under (2.19). Second, we apply the lemma

$$S_{[\mu}{}^{[3]} A_{\nu][3]} \equiv 0 ~, \tag{2.21}$$

for an arbitrary self-dual (or anti-self-dual) tensor $S_{[4]}$ (or $A_{[4]}$), to the tensors $f_{[4]}$ and $A_{[4]}$: $f_{[\mu}{}^{[3]} A_{\nu][3]} \doteq 0$. This verifies the vanishing of the last term of (2.18), as desired.

Eventually our supersymmetric self-dual system is characterized by the reduced $N = (1/8, 1)$ supersymmetry transformation rule

$$\delta_Q A_\mu{}^I = +i(\overline{\epsilon}_+ \gamma_\mu \lambda_-^I) ~, \tag{2.22a}$$

$$\delta_Q \lambda_-^I = -\tfrac{1}{4}(\gamma^{\mu\nu}\epsilon_-) F^{(+)I}_{\mu\nu} + (\gamma^\mu \epsilon_+) D_\mu B^I ~, \tag{2.22b}$$

$$\delta_Q B^I = +i(\overline{\epsilon}_- \lambda_-^I) ~, \tag{2.22c}$$

where $F^{(+)I}_{\mu\nu} \equiv P_{\mu\nu}{}^{\rho\sigma} F_{\rho\sigma}$ and $\epsilon_+$ subject to the condition (2.15). Now our supersymmetrized generalized self-duality conditions are (2.14), (2.17) and (2.15) all with the symbol $\stackrel{*}{=}$, consistent with supersymmetry (2.22). Eq. (2.17a) is nothing other than the generalized self-duality of $F_{\mu\nu}{}^I$, while (2.17b) deletes one degree of freedom out of two between the two fields $B$ and $C$.

We mention that the closure of supersymmetry after imposing the self-duality conditions (2.14) through (2.17) is also straightforward to confirm:

$$[\delta_Q(\epsilon_1), \delta_Q(\epsilon_2)] = \delta_P(\xi^\mu) ~,$$

$$\xi^\mu \stackrel{*}{=} +i(\overline{\epsilon}_{2+} \mathcal{P}\gamma^\mu \epsilon_{1-}) - i(\overline{\epsilon}_{1+} \mathcal{P}\gamma^\mu \epsilon_{2-})$$

$$\stackrel{*}{=} +i(\overline{\epsilon}_{2+} \gamma^\mu \epsilon_{1-}) - i(\overline{\epsilon}_{1+} \gamma^\mu \epsilon_{2-}) = +i(\overline{\epsilon}_2 \gamma^\mu \epsilon_1) ~. \tag{2.23}$$

In particular, the closure on $A_\mu{}^I$ needs a special care: Due to (2.15), we easily see that

$$-\tfrac{i}{4}(\overline{\epsilon}_{2+} \gamma_\mu \gamma^{\rho\sigma} \epsilon_{1-}) F^{(+)I}_{\rho\sigma} \stackrel{*}{=} -\tfrac{i}{4}(\overline{\epsilon}_{2+} [\gamma_\mu, \gamma^{\rho\sigma}] \epsilon_{1-}) F^{(+)I}_{\rho\sigma} = +i(\overline{\epsilon}_{2+} \gamma^\rho \epsilon_{1-}) F_{\rho\mu} ~. \tag{2.24}$$

Eventually, the commutator algebra (2.24) coincides formally with (2.3), but it is based on intricate spinorial properties associated with self-duality.

A special explanation is needed about our notation $N = (1/8, 1)$. This symbol shows the representations of the original supercharges in the $(\mathbf{8}_s, \mathbf{8}_c)$ of $SO(8)$ reduced to the $(\mathbf{1}, \mathbf{8})$ of $Spin(7)$. This is because of the embeddings $\mathbf{8}_s \to \mathbf{7} + \mathbf{1}$ and $\mathbf{8}_c \to \mathbf{8}$ under the holonomy reduction $SO(8) \to Spin(7)$. Since 7 components in the $\mathbf{8}_s$ are deleted, while $\mathbf{8}_c$ becomes the $\mathbf{8}$ of $Spin(7)$, the original supercharges $(\mathbf{8}_s, \mathbf{8}_c)$ become the $(\mathbf{1}, \mathbf{8})$ of $Spin(7)$. This counting is also obvious from the projection operators $\mathcal{Q}$ in (2.15) whose trace is seven, deleting seven components in $\epsilon_+$. The reason we use the quotient $1/8$ is that when both of the original supercharges are the $\mathbf{8}_s$ and $\mathbf{8}_c$, meaning $N = (1, 1)$ supersymmetry, we need a normalization factor 8, and the resulting supersymmetry is normalized to $N = (1/8, 1)$. Due to the asymmetry between left and right supercharges, this also means the 'chiral' nature of the system. Eventually in the self-dual multiplet $(A_\mu{}^I, \lambda_-{}^I, B^I)$ has $4 + 4$ physical degrees of freedom, because $\lambda_-{}^I$ has negative chirality with 8 components with 4 physical degrees of freedom, while $A_\mu{}^I$ is self-dual with 3 physical degrees of freedom. Accordingly, the parameters as well as the supersymmetry charges exist for *both* chiralities, like $\epsilon_+$ (1 component) and $\epsilon_-$ (8 components).

All of these can be also summarized in terms of supersymmetry algebra. The original $N = (1, 1)$ algebra

$$\{Q_\alpha, Q_\beta\} = (\gamma^\mu)_{\alpha\beta} P_\mu \tag{2.25}$$

is reduced into $N = (1/8, 1)$:

$$\{Q_{\alpha+}, Q_{\beta-}\} = (\mathcal{P}\gamma^\mu)_{\alpha+\beta-} P_\mu \ . \tag{2.26}$$

Due to the projector $\mathcal{P}$, the $\alpha$-index in (2.26) has effectively only one component corresponding to the **1** of $Spin(7)$, while $\beta = 1, 2, \cdots, 8$ for the **8** of $Spin(7)$. Eq. (2.26) is also equivalent to (A.9) in the Appendix, where the indices $\alpha$ and $\dot\beta$ are used in the latter instead of $\alpha+$ and $\beta-$ in the former.

Note also that if the generalized self-duality (2.17a) is substituted back into our lagrangian (2.1), the kinetic term of the $A_\mu$-field becomes a total divergence, as a topological invariant [3][5], as in self-dual vector theories in 4D [8][9]. Relevantly, the kinetic terms of $B$ and $C$ under (2.17b) cancel each other, explaining the opposite signs in their kinetic terms that looked unusual at first glance. As expected by supersymmetry, the kinetic term of $\lambda$ becomes also a total divergence under (2.14), as can be easily confirmed with the aid of (2.10) - (2.13). Additionally, the potential term in (2.1) also vanishes, resulting in a lagrangian completely vanishing modulo a total divergence, upon (2.14), (2.17a) and (2.17b).

The covariance of our generalized (anti)self-duality conditions (2.17a) and the generalized chirality condition (2.14) under supersymmetry provides a good confirmation of the consistency of our total system. The chirality in 8D collaborates with the generalized self-duality condition under an extra Killing spinor condition (2.15) for $N = (1/8, 1)$ supersymmetry.

## 3. Self-Dual $N = 1/8 + 7/8$ Supersymmetric Vector Multiplet in 7D

Once we have established the self-dual supersymmetric theory in 8D, a similar self-dual supersymmetric Yang-Mills theory in 7D can be also obtained by a simple dimensional reduction [15], with some additional constraints. The resulting theory will get the original $SO(7)$ Lorentz covariance now broken down to $G_2$. Interestingly, we can impose again a generalized self-duality on field strengths even though 7D is odd dimensional. Similarly to the case of 8D, the system has $N=2$ supersymmetry before imposing supersymmetric self-duality conditions. After imposing such self-duality conditions, $N=2$ supersymmetry is reduced to $N=1/8+7/8$ whose meaning will be clarified in the paragraph after (3.22).

The field content for $N=2$ supersymmetric Yang-Mills theory in Euclidean 7D is $(A_\mu{}^I, \lambda_A{}^I, B^I, C^I, A^I)$, where the new indices $A, B, \cdots = +, -$ are needed for doubling of 8-component Majorana spinor in 7D, corresponding to $N=2$ supersymmetries. The indices $+, -$ correspond to the positive and negative eigen-components of the Pauli matrix $\sigma_3$, as $(\sigma_3)_{++} = +1$, $(\sigma_3)_{--} = -1$. The new scalar field $A^I$ is from the 8-th component of the vector $A_\mu{}^I$ in 8D. We have in total $8+8$ physical degrees of freedom out of $A_\mu{}^I(5)$, $\lambda_A{}^I(2\times 4)$, $A^I(1)$, $B^I(1)$, $C^I(1)$. This explains the need of the additional index $A$ for the pseudo-Majorana [13] fermion $\lambda_A{}^I$. The charge conjugation matrix is symmetric in 7D [13], which can be chosen to be the Kronecker's delta: $C_{\alpha\beta} = \delta_{\alpha\beta}$. Relevantly, $\overline{\psi} = \psi^\dagger$, $\psi^* = \psi$, because both $A$ and $B$-matrices in [13] can be an unit matrix. Accordingly, we do not have to distinguish raising or lowering these spinorial indices $\alpha, \beta, \cdots$. The same is also true for the indices $A, B, \cdots$, as can be understood from dimensional reduction. The basic flipping and hermitian conjugation properties in 7D are $(\overline{\psi}_A \gamma^{[n]} \chi_B) = -(-1)^{n(n+1)/2} (\overline{\chi}_B \gamma^{[n]} \psi_A)$ and $(\overline{\psi}_A \gamma^{[n]} \chi_B)^\dagger = -(-1)^n (\overline{\psi}_A \gamma^{[n]} \chi_B)$.

Before giving our lagrangian, we give the basic rule of our simple dimensional reduction [15] for the $\gamma$-matrices. Using the symbol $\widehat{\phantom{x}}$ for all the 8D quantities and indices only in this section, we have

$$\widehat{C}_{\hat\alpha\hat\beta} = \delta_{\hat\alpha\hat\beta} = C_{\alpha\beta}\delta_{AB} = \delta_{\alpha\beta}\delta_{AB} \ ,$$

$$(\widehat{\gamma}_\mu)_{\hat\alpha\hat\beta} = (\gamma_\mu)_{\alpha\beta}(\sigma_2)_{AB} \ , \quad (\widehat{\gamma}_8)_{\hat\alpha\hat\beta} = \delta_{\alpha\beta}(\sigma_1)_{AB} \ , \quad (\widehat{\gamma}_9)_{\hat\alpha\hat\beta} = \delta_{\alpha\beta}(\sigma_3)_{AB} \ . \qquad (3.1)$$

Here the matrices $\sigma_1, \sigma_2, \sigma_3$ are the standard Pauli matrices. These ansatz are fixed by the requirement of the right Clifford algebra $\{\gamma_\mu, \gamma_\nu\} = +2\delta_{\mu\nu}$, the antisymmetry $(\gamma_\mu)^T = -\gamma_\mu$, and the hermiticity $(\gamma_\mu)^\dagger = \gamma_\mu$ in 7D [13]. Accordingly, we have the field assignments

$$\widehat{A}_{\hat\mu} = \begin{cases} \widehat{A}_\mu{}^I = A_\mu{}^I \ , \\ \widehat{A}_8{}^I = A^I \ , \end{cases} \qquad (3.2\text{a})$$

$$\widehat{\lambda}_{\hat\alpha} = \lambda_{\alpha A} \ , \quad \widehat{B}^I = B^I \ , \quad \widehat{C}^I = C^I \ . \qquad (3.2\text{b})$$

We have also to comply with the above-mentioned symmetry and hermiticity for $(\overline{\psi}_A \gamma^{[n]} \chi_B)$ in 7D [13]. These properties explain the needs of an extra antisymmetric matrix $\sigma_2$ and the imaginary unit in the parameter of translation $\xi^\mu \equiv +i(\overline{\epsilon}_2 \sigma_2 \gamma^\mu \epsilon_1)$, such that it is antisymmetric under the two supersymmetry parameters $\epsilon_1 \leftrightarrow \epsilon_2$.

Applying these rules to our previous 8D system, we get our lagrangian in 7D as

$$\begin{aligned}\mathcal{L}_{7D} = &-\tfrac{1}{4}(F_{\mu\nu}{}^I)^2 - i(\overline{\lambda}^I \sigma_2 \gamma^\mu D_\mu \lambda^I) + \tfrac{1}{2}(D_\mu B^I)^2 - \tfrac{1}{2}(D_\mu C^I)^2 - \tfrac{1}{2}(D_\mu A^I)^2 \\ &- if^{IJK} B^I (\overline{\lambda}^J \lambda^K) + if^{IJK} C^I (\overline{\lambda}^J \sigma_3 \lambda^K) + if^{IJK} A^I (\overline{\lambda}^J \sigma_1 \lambda^K) \\ &+ \tfrac{1}{2}(f^{IJK} B^J C^K)^2 + \tfrac{1}{2}(f^{IJK} A^J B^K)^2 - \tfrac{1}{2}(f^{IJK} C^J A^K)^2 \;, \end{aligned} \quad (3.3)$$

whose action $I_{7D} \equiv \int d^7 x\, \mathcal{L}_{7D}$ is invariant under $N=2$ supersymmetry

$$\begin{aligned} \delta_Q A_\mu{}^I &= +i(\overline{\epsilon}\sigma_2 \gamma_\mu \lambda^I) \;, \\ \delta_Q \lambda_A{}^I &= -\tfrac{1}{4}\gamma^{\mu\nu}\epsilon_A F_{\mu\nu}{}^I + \tfrac{1}{2}(\sigma_2 \gamma^\mu \epsilon)_A D_\mu B^I - \tfrac{i}{2}(\sigma_1 \gamma^\mu \epsilon)_A D_\mu C^I + \tfrac{i}{2}(\sigma_3 \gamma^\mu \epsilon)_A D_\mu A^I \\ &\quad - \tfrac{1}{2}f^{IJK}(\sigma_3 \epsilon)_A B^J C^K + \tfrac{1}{2}f^{IJK}(\sigma_1 \epsilon)_A A^J B^K + \tfrac{i}{2}f^{IJK}(\sigma_2 \epsilon)_A A^J C^K \;, \\ \delta_Q B^I &= +i(\overline{\epsilon}\lambda^I) \;, \quad \delta_Q C^I = +i(\overline{\epsilon}\sigma_3 \lambda^I) \;, \quad \delta_Q A^I = +i(\overline{\epsilon}\sigma_1 \lambda^I) \;. \end{aligned} \quad (3.4)$$

Since we do not have to distinguish raising and lowering the indices $A$, $B$, $\cdots$, the multiplication by the Pauli matrices is easy, e.g., $(\overline{\lambda}^I \sigma_2 \gamma^\mu D_\mu \lambda^I) \equiv \lambda_{\alpha A}{}^I (\sigma_2)_{AB}(\gamma^\mu)_{\alpha\beta} D_\mu \lambda_{\beta B}{}^I$. As in the 8D case, we can easily confirm the commutator algebra

$$[\delta_Q(\epsilon_1), \delta_Q(\epsilon_2)] = \delta_P(\xi^\mu) \;, \quad \xi^\mu \equiv i(\overline{\epsilon}_2 \sigma_2 \gamma^\mu \epsilon_1) \equiv i\epsilon_{2\alpha A}(\sigma_2)_{AB}(\gamma^\mu)_{\alpha\beta} \epsilon_{1\beta B} \;. \quad (3.5)$$

If one is not familiar with the usage of the antisymmetric matrix $\sigma_2$, one is referred to eq. (2.2) in ref. [16] on 6D supergravity, where supertorsion has the component $T_{\underline{\alpha\beta}}{}^c = i(\sigma^c)_{\alpha\beta}\epsilon_{ij}$ with the antisymmetric metric $\epsilon_{ij}$ of $Sp(1)$ group. Note that the $\gamma$-matrix in 6D is antisymmetric, and that is the reason why we need the extra tensor $\epsilon_{ij}$. Due to the similarity in $\gamma$-matrix structures between 7D and 6D, it is natural that we have such an antisymmetric matrix in (3.5).

We now look into supersymmetric generalized self-duality in 7D. There is a subtlety associated with the compatibility of simple dimensional reduction [15] from 8D and generalized self-duality within 7D. Upon the dimensional reduction from 8D, we expect that eqs. (2.17) and (2.14) yield conditions like $F_{\mu\nu}{}^I \stackrel{*}{=} (1/2)f_{\mu\nu}{}^{\rho\sigma}F_{\rho\sigma}{}^I$ or $B^I \stackrel{*}{=} -C^I$, or a condition corresponding to $\widehat{P}\widehat{\lambda}^I \stackrel{*}{=} 0$, etc. However, we soon find that this is not enough for supersymmetry. For example, we notice that the scalar field $A^I$ is to vanish, because it will generate a new gradient term of $A$ in 7D out of $\widehat{F}_{\hat{\mu}\hat{\nu}}{}^I \stackrel{*}{=} (1/2)\widehat{f}_{\hat{\mu}\hat{\nu}}{}^{\hat{\rho}\hat{\sigma}} \widehat{F}_{\hat{\rho}\hat{\sigma}}{}^I$. Once $A$ vanishes, its supersymmetric transformation (3.4) tells that the combination $(\overline{\epsilon}\sigma_1 \lambda)$ should also vanish. However, we do not get such a condition directly by a simple dimensional reduction [15] of the conditions on fields (2.14) and (2.17) in 8D, or on the constraint (2.15) on the parameter $\widehat{\epsilon}$.

After studying the mutual consistency of these conditions, we arrive at the following set of constraints on fields:

$$F_{\mu\nu}{}^I \stackrel{*}{=} \tfrac{1}{2}\phi_{\mu\nu}{}^{\rho\sigma} F_{\rho\sigma}{}^I \;, \quad (3.6a)$$

$$\lambda_+{}^I \overset{*}{=} 0 \ , \tag{3.6b}$$

$$\mathcal{P}\lambda_-{}^I \overset{*}{=} 0 \ , \tag{3.6c}$$

$$B^I \overset{*}{=} -C^I \ , \tag{3.6d}$$

$$A^I \overset{*}{=} 0 \ , \tag{3.6e}$$

and constraints on the parameter of supersymmetry (Killing spinor conditions) [3]:

$$\gamma_{(+)}^{\mu\nu} \epsilon_+ \equiv P^{\mu\nu\rho\sigma} \gamma_{\rho\sigma} \epsilon_+ \overset{*}{=} 0 \ , \tag{3.7a}$$

$$\mathcal{N}\epsilon_+ \overset{*}{=} 0 \ , \tag{3.7b}$$

$$\mathcal{P}\epsilon_- \overset{*}{=} 0 \ . \tag{3.7c}$$

Eq. (3.7b) is a necessary condition of (3.7a). The $\phi_{\mu\nu\rho\sigma}$ are given in (2.6) with $\phi^{[4]} \equiv (1/3!)\, \epsilon^{[4][3]} \psi_{[3]}$, to be distinguished from $\widehat{f}_{\hat\mu\hat\nu\hat\rho\hat\sigma}$ due to the absence of components like $\phi_{\mu\nu\rho 8}$. The subscripts $\pm$ are for the components $A = \pm$, and the symbols $P_{\mu\nu}{}^{\rho\sigma}$, $\mathcal{P}$ and $\mathcal{N}$ are projectors defined by [6][3][5]

$$P_{\mu\nu}{}^{\rho\sigma} \equiv +\tfrac{2}{3}\big(\delta_{[\mu}{}^\rho \delta_{\nu]}{}^\sigma + \tfrac{1}{4}\phi_{\mu\nu}{}^{\rho\sigma}\big) \ , \quad N_{\mu\nu}{}^{\rho\sigma} \equiv +\tfrac{1}{3}\big(\delta_{[\mu}{}^\rho \delta_{\nu]}{}^\sigma - \tfrac{1}{2}\phi_{\mu\nu}{}^{\rho\sigma}\big) \ , \tag{3.8a}$$

$$\mathcal{P} \equiv +\tfrac{1}{8}(I + \Psi) \ , \quad \mathcal{N} \equiv +\tfrac{7}{8}\big(I - \tfrac{1}{7}\Psi\big) \ , \quad \Psi \equiv +\tfrac{1}{4!}\phi^{[4]}\gamma_{[4]} = +\tfrac{i}{3!}\psi^{[3]}\gamma_{[3]} \ . \tag{3.8b}$$

The important consistency is among the conditions (3.6) and (3.7) and supersymmetry (3.4), as was done in 8D. The simplest confirmation is the supersymmetric transformation of (3.6d) which vanishes upon (3.6b). Under (3.6a), the variation of (3.6b) under supersymmetry yields five different terms all of which vanish upon the use of other conditions (3.6).

The non-trivial ones are the supersymmetric variations of (3.6a), (3.6c) and (3.6e). The first one is

$$0 \overset{?}{=} \delta_Q\big(F_{\mu\nu}{}^I - \tfrac{1}{2}\phi_{\mu\nu}{}^{\rho\sigma} F_{\rho\sigma}{}^I\big) \overset{*}{=} -6 N_{\mu\nu}{}^{\rho\sigma}(\gamma_\rho)_{\alpha\beta}(\epsilon_{\alpha+} D_\sigma \lambda_{\beta-}{}^I)$$

$$\overset{*}{=} -6i N_{\mu\nu}{}^{\rho\sigma}\big(\psi_{\rho\alpha\beta} + \delta_{\rho\alpha}\delta_{\beta 8} - \delta_{\rho\beta}\delta_{\alpha 8}\big)(\epsilon_{\alpha+} D_\sigma \lambda_{\beta-}{}^I) \ , \tag{3.9}$$

where use is made of the expression of the $\gamma$-matrix in terms of $\psi$ [6]:

$$(\gamma_\mu)_{\alpha\beta} = i(\psi_{\mu\alpha\beta} + \delta_{\mu\alpha}\delta_{\beta 8} - \delta_{\mu\beta}\delta_{\alpha 8}) \ . \tag{3.10}$$

Our next step is to see what components among the $\epsilon_+$'s and $\lambda_-$'s have nonzero values. Note also that $\mathcal{P}$ and $\mathcal{N}$ can be expressed more explicitly like (3.10), as

$$\mathcal{P}_{\alpha\beta} = \delta_{\alpha 8}\delta_{\beta 8} \ , \quad \mathcal{N}_{\alpha\beta} = \delta_{\alpha\beta} - \delta_{\alpha 8}\delta_{\beta 8} \ . \tag{3.11}$$

These with (3.7b), (3.7c) and (3.6c) imply that

$$\epsilon_{\alpha+} \overset{*}{=} 0 \quad (\text{for } \alpha = 1, 2, \cdots, 7) \ , \qquad \epsilon_{8-} \overset{*}{=} 0 \ , \qquad \lambda_{8-}{}^I \overset{*}{=} 0 \ . \tag{3.12}$$

In other words, there is only one non-zero component $\epsilon_{8+}$ in $\epsilon_+$, only seven non-zero components in $\epsilon_-$, and only seven non-zero components $\lambda_{\alpha-}{}^I$ ($\alpha = 1, 2, \cdots, 7$) in $\lambda_-{}^I$. This is also related to our notation $N = 1/8 + 7/8$, which we will be back shortly. Considering these, we simplify (3.9) as

$$0 \overset{?}{=} -6i N_{\mu\nu}{}^{\rho\sigma}(\epsilon_{8+} D_\rho \lambda_{\sigma-}{}^I) \ . \tag{3.13}$$

We can finally show that this vanishes by the use of the $\lambda$-field equation

$$\gamma^\mu D_\mu \lambda_-{}^I \doteq 0 \ . \tag{3.14}$$

under (3.6d) and (3.6e). In fact, (3.14) yields with the aid of (3.10),

$$i\psi^\mu{}_{\alpha\beta} D_\mu \lambda_{\beta-}{}^I - \delta_{\alpha 8} D^\mu \lambda_{\mu-}{}^I \overset{*}{=} 0 \quad \Longrightarrow \quad \psi_\mu{}^{\rho\sigma} \epsilon_{8+} D_\rho \lambda_{\sigma-}{}^I \overset{*}{=} 0 \ . \tag{3.15}$$

Multiplying this by $\psi_{\tau\lambda}{}^\mu$, we get

$$N_{\mu\nu}{}^{\rho\sigma}(\epsilon_{8+} D_\rho \lambda_{\sigma-}{}^I) \overset{*}{=} 0 \ , \tag{3.16}$$

due to the identity $\psi_{\mu\nu\tau}\psi^{\tau\rho\sigma} = +6 N_{\mu\nu}{}^{\rho\sigma}$. Eq. (3.16) means exactly the vanishing of the last side of (3.13), and therefore (3.9) is now confirmed.

The next one is the variation of (3.6e):

$$0 \overset{?}{=} \delta_Q A^I = +i(\overline{\epsilon}\sigma_1 \lambda^I) = +i(\overline{\epsilon}_+ \lambda_-{}^I) + i(\overline{\epsilon}_- \lambda_+{}^I)$$
$$\overset{*}{=} +i(\overline{\epsilon}_+ \lambda_-{}^I) = +i(\overline{\epsilon}_+ \mathcal{P}\lambda_-{}^I) = 0 \qquad (Q.E.D.) \ , \tag{3.17}$$

by help of (3.7b) and (3.6c). This justifies the necessity of the condition (3.6c).

The last non-trivial supersymmetric variation is $\delta_Q(\mathcal{P}\lambda_-{}^I)$:

$$0 \overset{?}{=} \delta_Q(\mathcal{P}\lambda_-{}^I) = \mathcal{P}\delta_Q \lambda_-{}^I \overset{*}{=} +\tfrac{i}{8}\gamma^\mu \epsilon_+ D_\mu B^I - \tfrac{1}{16}\psi_\mu{}^{\rho\sigma}\gamma^{(-)}_{\rho\sigma}\epsilon_+ D_\mu B^I - \tfrac{i}{48}\phi_\mu{}^{\tau\rho\sigma}\gamma_\tau \gamma^{(-)}_{\rho\sigma}\epsilon_+ D_\mu B^I$$
$$\overset{*}{=} -\tfrac{1}{16}\Big(6i\gamma^\mu \epsilon_+ + \psi_\mu{}^{\rho\sigma}\gamma_{\rho\sigma}\epsilon_+\Big) D_\mu B^I$$
$$\overset{*}{=} -\tfrac{3i}{8}\Big(i\psi_{\mu\alpha\beta} + i\delta_{\mu\alpha}\delta_{\beta 8} - i\delta_{\mu\beta}\delta_{\alpha 8}\Big)\epsilon_{\beta+} - \tfrac{1}{16}\psi_\mu{}^{\rho\sigma}\Big(\phi_{\rho\sigma\alpha\beta} + \psi_{\rho\sigma\alpha}\delta_{\beta 8} - \psi_{\rho\sigma\alpha}\delta_{\beta 8} + 2\delta_{\rho\alpha}\delta_{\sigma\beta}\Big)\epsilon_{\beta+}$$
$$\overset{*}{=} +\tfrac{3}{8}\delta_{\mu\alpha}\epsilon_{8+} - \tfrac{3}{8}\delta_{\mu\alpha}\epsilon_{8+} = 0 \qquad (Q.E.D.) \ , \tag{3.18}$$

where we have used (3.6), (3.7) and (3.12), with other basic facts such as $\psi_{\mu\nu 8} = 0$, etc., as well as the non-trivial identities

$$\phi_{\mu\nu}{}^{\rho\sigma}\gamma^{(-)}_{\rho\sigma} = -4\gamma^{(-)}_{\mu\nu} \ , \qquad \Psi\gamma^\mu = +\tfrac{i}{2}\psi_\mu{}^{\rho\sigma}\gamma_{\rho\sigma} - \tfrac{1}{6}\phi_\mu{}^{[3]}\gamma_{[3]} \ , \qquad [\mathcal{P}, \gamma^{\mu\nu}_{(+)}] = 0 \ , \tag{3.19}$$

with $\gamma^{(-)}_{\mu\nu} \equiv N_{\mu\nu}{}^{\rho\sigma}\gamma_{\rho\sigma}$, and the expression [6]

$$(\gamma_{\mu\nu})_{\alpha\beta} = \phi_{\mu\nu\alpha\beta} + (\psi_{\mu\nu\alpha}\delta_{\beta 8} - \psi_{\mu\nu\beta}\delta_{\alpha 8}) + (\delta_{\mu\alpha}\delta_{\nu\beta} - \delta_{\mu\beta}\delta_{\nu\alpha}) \ . \tag{3.20}$$

As an additional consistency check, we can study if all the field equations from (3.3) are consistent with (3.6). First, for the index $A = -$ the $\lambda$-field equation is easily seen to be satisfied under (3.6). Second, for the index $A = +$, it dictates the field equation for $\lambda_-{}^I$, which has been already given in (3.14). Third, the $A$-field equation vanishes almost trivially under (3.6). Fourth, the sum of $B$ and $C$-field equations vanish desirably under (3.6). Fifth, the difference of these equations yields the field equation for the active component $2B^I \stackrel{*}{=} B^I - C^I$. Sixth, the $A_\mu{}^I$-field equation becomes simply $D_\nu F_\mu{}^{\nu I} \stackrel{*}{=} 0$ under (3.6). However, this can be skipped, because it is a necessary condition of the generalized duality (3.6a). As for the compatibility between $\mathcal{P}\lambda_-{}^I \stackrel{*}{=} 0$ (3.6c) and the Dirac equation (3.14), we have already seen that the projector $\mathcal{P}$ makes only the 8-th component of $\lambda_-{}^I$ vanish, with no problem with (3.14).

Therefore, our field equations for our $N = 1/8 + 7/8$ supersymmetric self-dual Yang-Mills theory are

$$\gamma^\mu D_\mu \lambda_-{}^I \stackrel{*}{=} 0 \ , \tag{3.21a}$$

$$D_\mu^2 B^I \stackrel{*}{=} 0 \ . \tag{3.21b}$$

Needless to say, we have also the supersymmetric self-duality condition (3.6) with the constrained parameter (3.7). This provides an additional confirmation of the consistency of our constraints (3.6) under (3.7).

Let now us study the closure of supersymmetry in our self-dual system. After imposing (3.6) and (3.7), the closure $[\delta_Q(\epsilon_1), \delta_Q(\epsilon_2)] = \delta_P(\xi^\mu)$ now has the parameter for the translation $P_\mu$:

$$\xi_\mu = \epsilon_{2\alpha+}(\gamma_\mu)_{\alpha\beta}\epsilon_{1\beta-} - \epsilon_{1\alpha+}(\gamma_\mu)_{\alpha\beta}\epsilon_{2\beta-} = +\epsilon_{18+}\epsilon_{2\mu-} - \epsilon_{28+}\epsilon_{1\mu-} \ , \tag{3.22}$$

where we used (3.10) and (3.12). In particular, the index $8$ as in $\epsilon_{18+}$ is for the eighth-component of $\epsilon_{1\alpha+}$ which is now in the $\mathbf{1}$ of $G_2$, while $\mu = 1, 2, \cdots, 7$ as the $\mathbf{7}$ of $G_2$. As this non-vanishing expression shows explicitly, there is a translation generated out of two supersymmetries, even within the reduced $G_2$ Lorentz covariance! Note that we have shown this as a classical system, instead of any quantized field theory. Of course, the original $SO(7)$ Lorentz covariance is lost and reduced into $G_2$, but this is exactly how we want to formulate the self-dual supersymmetric Yang-Mills theory in 7D.

As in the previous 8D case, the caveat here is about the expression $N = 1/8 + 7/8$ supersymmetry. This is equivalent to a more precise expression $N = (1/8)_+ + (7/8)_-$, meaning that the original $N = 2$, i.e., $N = 1_+ + 1_-$ consisting of two supersymmetries with $+$ and $-$ eigen-components of $\sigma_3$ are now reduced to $N = (1/8)_+ + (7/8)_-$. In other words, supersymmetry now undergoes peculiar constraints such as (3.12) that eliminates 7 components in $Q_{\alpha+}$, and one component in $Q_{\alpha-}$. Eventually, as (3.22) clearly shows, the effective spinor charges are now $Q_{8+}$ and $Q_{\mu-}$ ($\mu = 1, 2, \cdots, 7$), while $Q_{\mu+}$ and $Q_{8-}$ have been truncated. In terms of the Lorentz covariance breaking, eqs. (3.6c), (3.7b) and (3.7c) imply that under $SO(7) \to G_2$, $\lambda_-{}^I$ is reduced like $\mathbf{8} \to \mathbf{7}$, while $\epsilon_+$ is reduced as $\mathbf{8} \to \mathbf{1}$, and $\epsilon_-$ as $\mathbf{8} \to \mathbf{7}$. In terms of supersymmetry algebra, the original $N = 2$ algebra

$$\{Q_{\alpha A}, Q_{\beta B}\} = +(\gamma_\nu)_{\alpha\beta}(\sigma_2)_{AB} P_\nu = +(\psi_{\nu\alpha\beta} + \delta_{\nu\alpha}\delta_{\beta 8} - \delta_{\nu\beta}\delta_{\alpha 8})(\sigma_2)_{AB} P_\nu \quad , \quad (3.23)$$

is now reduced into $N = 1/8 + 7/8$ as

$$\{Q_{8+}, Q_{\mu-}\} = +i P_\mu \quad , \quad (3.24)$$

where $Q_{8+}$ is the $\mathbf{1}$ of $G_2$, while $Q_{\mu-}$ is the $\mathbf{7}$ of $G_2$.

We mention that the conditions (3.6b) and the constraint (3.7c) do not come from 8D *via* a simple dimensional reduction [15]. Even though this looks strange at first sight, it is natural from the viewpoint that simple dimensional reductions [15] may not respect the subtlety of the embedding of $G_2$ symmetry into the full $SO(7)$ Lorentz covariance in 7D.

Note that the previous chirality in 8D needed for consistency with generalized self-duality is now converted into the eigenvectors for the projectors $(I \pm \sigma_3)/2$ in 7D which is for 'internal' space. In other words, the superpartner condition for generalized self-duality in 7D requires the 'chirality' for the internal space, because odd dimensional space like 7D can not define chirality. Due to this internal chirality condition, the resulting supersymmetry in 7D becomes effectively $N = 1/8 + 7/8$ reduced from the initial $N = 2$ in (3.3). It is not only the Lorentz symmetry reduced from $SO(7)$ into $G_2$, but also the number of supersymmetries from $N = 2$ to $N = 1/8 + 7/8$ upon the requirement of generalized self-duality in 7D.

## 4. Concluding Remarks

In this paper, we have presented a vector multiplet with generalized self-duality with $N = (1/8, 1)$ supersymmetry in 8D, and a similar theory with $N = 1/8 + 7/8$ supersymmetry in 7D. This formulation is a generalization of the self-dual supersymmetric vector multiplet in 4D [8][9] into higher dimensions $4 \leq D \leq 8$. Even though we dealt only with the case of generalized self-dualities in 8D and 7D, we can repeat similar formulation for 5D and 6D.

By help of the special 4-th rank antisymmetric $SO(7)$-invariants $f_{\mu\nu\rho\sigma}$ defined in terms of octonion structure constants $\psi_{\mu\nu\rho}$ ($\mu, \nu, \rho = 1, 2, \cdots, 7$), we are able to define the generalized self-duality for the basic $N = (1,1)$ vector multiplet in 8D with non-trivial non-Abelian interactions. The new ingredient in our formulation is the introduction of constraints on the supersymmetry parameter in 8D, in such a way that the chirality condition on the fermions will not delete all the degrees of freedom, consistently with $N = (1/8, 1)$ supersymmetry. Due to the chirality condition required by self-duality, supersymmetry is also reduced from $N = (1,1)$ into $N = (1/8, 1)$. Based on this result, we have also obtained an $N = 2$ supersymmetric vector multiplet in 7D by a simple dimensional reduction [15], and we have obtained certain $N = 1/8 + 7/8$ supersymmetric self-duality conditions in 7D, with extra constraints that are not from 8D. The expressions such as $N = (1/8, 1)$ in 8D or $N = 1/8 + 7/8$ in 7D need some caution, because they respectively mean that the original $N = (1,1)$ or $N = 2$ supersymmetries are constrained under such peculiar constraints (2.15) or (3.12). This subtlety also reflects the important fact that in these self-dual supersymmetric models in 8D or 7D, the reduction of supersymmetries is much more sophisticated than other analogous cases without self-dualities. We have seen that the octonion structure constants, generalized self-duality, $G_2$ holonomy and generalized chirality are all closely related under $N = (1/8, 1)$ supersymmetry in 8D, and $N = 1/8 + 7/8$ supersymmetry in 7D.

We have seen both in 8D and 7D that after imposing the supersymmetric self-duality conditions, the original lagrangian completely vanish up to a total divergence. This aspect is quite common in self-dual theories, such as supersymmetric self-dual Yang-Mills theories in $D = 2+2$ [8][9]. In fact, this was the first difficulty for action formulation for supersymmetric self-dual Yang-Mills theories in $D = 2+2$, as has been emphasized in [9]. A similar situation is found in Type IIB supergravity theory in 10D, where the kinetic term of a self-dual 4-th rank potential $A_{\mu_1\cdots\mu_4}$ vanishes identically, after imposing the self-duality condition on its field strength: $F_{[5]} \stackrel{*}{=} (1/5!)\epsilon_{[5]}{}^{[5]'} F_{[5]'}$, because $(F_{[5]})^2 \stackrel{*}{=} 0$. Since we are dealing with supersymmetric system, it is quite natural to have all the lagrangian terms vanishes, once supersymmetric self-duality conditions are imposed on all the fields, starting with the bosonic field strength of vector fields.

In order to avoid confusion or misunderstanding, we stress the following: The important aspect of the present result is that we have shown it is possible to formulate self-dual supersymmetric Yang-Mills theories in 8D and 7D at the classical level, instead of quantum level, with the original holonomy $SO(8)$ or $SO(7)$ reduced to $Spin(7)$ or $G_2$, respectively. Of course, the original Lorentz symmetries in these theories are lost, but we can still manage the whole systems, working under appropriate supersymmetric constraints with the right projection operators. Note also that we have managed to formulate these theories, consistently with supersymmetries. There have been some works on self-dual theories and topological field theories based on quantized systems, such as those in [10][17]. However, our main result in this paper is that we can formulate self-dual theories at the classical level, before quantization.

We also stress that our self-dual theory in 8D can be also regarded as the global supersymmetric version of self-dual supergravity in 8D given in [14]. In fact, in the next Appendix we have performed the superspace reformulation of our component result for self-dual supersymmetric Yang-Mills theory in 8D, in order to verify the nontrivial features of reduced Lorentz symmetry $Spin(7)$. As has been also clarified in [14] for self-dual supergravity in 8D, there is nothing wrong with imposing self-duality condition that reduces the original holonomy $SO(8)$ into $Spin(7)$, as has been explicitly demonstrated in superspace [14], consistently with local supersymmetry and general coordinate covariances as well.

There are some expected as well as non-trivial ingredients in our formulations. For example, we know that the bosonic field strength $F_{\mu\nu}$ can undergo generalized self-duality in 8D [5][3], anticipated from 4D self-duality [8][9]. Moreover, recent studies on M-theory indicate the existence of such supersymmetric theories in 7D with $G_2$ holonomy in 11D supergravity compactified into 4D, in particular with certain Killing spinor conditions [3][6][5]. Such $G_2$ holonomy is in the mathematical sequence of holonomies of $SO(7) \supset G_2 \supset SU(3) \supset SU(2)$ in $D = 8, 7, 6$ and $4$, respectively. However, the constrained parameters of supersymmetry by (2.15) or (3.7) are non-trivial, because no conventional supermultiplet has such constraints.

Since our theory can be also dimensionally reduced into lower dimensions $D < 8$, our theory in 8D can serve as an important starting point of these self-dual supersymmetric Yang-Mills theories in $7 \geq D \geq 4$. In particular, as we have seen, this includes the most interesting case of $G_2$ covariance in 7D [3], which is more relevant to the extra space in M-theory compactifications from 11D into 4D. Note also that supersymmetric generalized self-duality theory can serve as a underlying theory even for the self-dual theory in 4D [7][8][5][9],[10] which in turn had been supposed to be the origin of all the integrable models in dimensions in $3 \geq D \geq 1$ [7].

In our self-dual supersymmetric theories, we saw that the substitutions of self-duality conditions (2.14), (2.17) or (3.6) back to the initial lagrangians (2.1) or (3.3) make these lagrangians vanish modulo a total divergence. This is similar to Type IIB theory in 10D, related to the self-duality of the fifth-rank tensor $F_{[5]}$. The difference is that in the present case, all the fields will lose their initial degrees of freedom, because not only the $F_{[2]}^2$-term but all the kinetic terms vanish up to a total divergence.

We stress that our result in this paper provides the first explicit formulation of supersymmetric generalized self-dualities in 8D and 7D, to our knowledge. For example, the lagrangian for supersymmetric Yang-Mills multiplet (2.1) in Euclidean 8D has been known for some time [11], before imposing generalized self-duality. Or non-supersymmetric generalized self-duality theories in higher-dimensions have been known for some time [3][5][4][11][10]. However, the important point is that we have studied the mutual consistency among generalized self-dualities, reduced $G_2$ or $Spin(7)$ symmetries [2][3][4], octonion algebra [6], Clifford algebra, chirality/parity, and supersymmetry, with supersymmetry parameters with peculiar constraints (2.15) and (3.7), not in a conjectural or abstract way, but in an explicit, detailed, self-contained and closed form.

---

[10]This statement is valid up to the signature of space-time. As a matter of fact, there is a non-compact $G_2$ group as a subgroup of $SO(4,3)$ with the signature $(++++---)$, which might be more relevant in this context. We acknowledge M. Günaydin about this [18].

The result of this paper shows another important aspect of supersymmetry. In the past, we have been more accustomed to supersymmetry or superspace with manifest Lorentz covariance in any space-time dimensions. However, our result here shows that there are many still unknown supersymmetric theories whose Lorentz covariance is not manifest, or partially broken, such as $SO(8) \to SO(7)$ in 8D, or $SO(7) \to G_2$ in 7D. In fact, more supersymmetric theories with no manifest Lorentz covariance have been formulated nowadays, such as supersymmetric Yang-Mills theory in $D \geq 12$ with no manifest Lorentz covariance [19][20], or massive supergravity formulation with Killing vector in 11D [21], or teleparallel superspace formulation [22] with no manifest Lorentz connection. All of these results indicate the importance of no manifest Lorentz covariance, with more significance yet to be discovered in the future. Even though we have dealt in this paper only with global supersymmetry, our result provides with the foundation of locally supersymmetric cases, such as self-dual supergravity in 8D in terms of topological formulation [23] or in 7D with $G_2$ holonomy [24].

We are grateful to the referee of this paper who gave many important suggestions.

**Appendix: Superspace for Self-Dual Supersymmetric Yang-Mills in 8D**

This appendix is devoted to the superspace re-formulation of our self-dual supersymmetric Yang-Mills in 8D. This appendix will of great help for understanding our commutator algebra, and the total consistency of highly nontrivial self-dual system in 8D with the reduced Lorentz covariance $Spin(7)$. From a superspace viewpoint, the formulation below is nothing but an analog with global supersymmetry of supergravity in superspace with the reduced holonomy $Spin(7)$ [14].

Our superspace formulation for $N = (1,1)$ in 8D is based on the $F$- and $T$-Bianchi identities (BIds)

$$\nabla_{[A} F_{BC)}{}^I - T_{[AB|}{}^D F_{D|C)}{}^I \equiv 0 \ , \tag{A.1a}$$

$$\nabla_{[A} T_{BC)}{}^D - T_{[AB|}{}^E T_{E|C)}{}^D - \tfrac{1}{2} R_{[AB|e}{}^f (\mathcal{M}_f{}^e)_{|C)}{}^D \equiv 0 \ , \tag{A.1b}$$

where the superfield strength $F$ is defined by

$$F_{AB}{}^I \equiv \nabla_{[A} A_{B)}{}^I - T_{AB}{}^C A_C{}^I + f^{IJK} A_A{}^J A_B{}^K \ , \tag{A.2}$$

in terms of the superpotential $A_A{}^I$. Only in this Appendix for superspace, we use the different index convention, such as $A, B, \cdots$ for local Lorentz indices in superspace both for bosonic and fermionic indices: $A \equiv (a, \underline{\alpha})$, $B \equiv (b, \underline{\beta})$, $\cdots$, where $a, b, \cdots = 1, 2, \cdots, 8$ are for the bosonic, and $\underline{\alpha}, \underline{\beta}, \cdots = 1, 2, \cdots, 16$ are for the fermionic indices. The superspace convention is also for the antisymmetrization, such as $X_{[AB)} \equiv X_{AB} - (-1)^{AB} X_{BA}$ with *no* factor $1/2$, where the factor $(-1)^{AB}$ counts the relative Grassmann parity of the indices $A$ and $B$, as usual.

The field content of our system *before* imposing any self-duality conditions is $(A_a{}^I, \lambda_{\underline{\alpha}}{}^I, B^I, C^I)$. As usual, for a supersymmetric Yang-Mills theory, we set up the constraints, satisfying the BIds mainly at dimensions $0 \le d \le 1$:

$$\begin{aligned}
& T_{\underline{\alpha}\underline{\beta}}{}^c = +i(\gamma^c)_{\underline{\alpha}\underline{\beta}} \ , \quad T_{\underline{\alpha}\underline{\beta}}{}^{\underline{\gamma}} = 0 \ , \quad T_{\alpha b}{}^D = 0 \ , \quad T_{ab}{}^D = 0 \ , \\
& F_{\underline{\alpha}\underline{\beta}}{}^I = -i\delta_{\underline{\alpha}\underline{\beta}} B^I + i(\gamma_9)_{\underline{\alpha}\underline{\beta}} C^I \ , \\
& F_{\underline{\alpha} b}{}^I = -i(\gamma_b)_{\underline{\alpha}\underline{\beta}} \lambda_{\underline{\beta}}{}^I \equiv -i(\gamma_b \lambda^I)_{\underline{\alpha}} \ , \\
& \nabla_{\underline{\alpha}} \lambda_{\underline{\beta}}{}^I = -\tfrac{1}{4}(\gamma^{ab})_{\underline{\alpha}\underline{\beta}} F_{ab}{}^I - \tfrac{1}{2}(\gamma^a)_{\underline{\alpha}\underline{\beta}} \nabla_a B^I + \tfrac{1}{2}(\gamma_9 \gamma^a)_{\underline{\alpha}\underline{\beta}} \nabla_a C^I + f^{IJK}(\gamma_9)_{\underline{\alpha}\underline{\beta}} B^J C^K \ , \\
& \nabla_{\underline{\alpha}} B^I = -i \lambda_{\underline{\alpha}}{}^I \ , \\
& \nabla_{\underline{\alpha}} C^I = -i(\gamma_9 \lambda^I)_{\underline{\alpha}} \ ,
\end{aligned} \tag{A.3}$$

which are consistent with our component transformation rule (2.2).

As usual in superspace, the superfield equations for all the fields are obtained from BIds at $d \ge 1$, as

$$\begin{aligned}
& i(\slashed{\nabla} \lambda^I)_{\underline{\alpha}} + i f^{IJK} \lambda_{\underline{\alpha}}{}^J B^K - i f^{IJK}(\gamma_9 \lambda^I)_{\underline{\alpha}} C^K \doteq 0 \ , \\
& \nabla_b F_a{}^{bI} - i f^{IJK}(\overline{\lambda}^J \gamma_a \lambda^K) - f^{IJK} B^J \nabla_a B^K + f^{IJK} C^J \nabla_a C^K \doteq 0 \ , \\
& \nabla_a^2 B^I + i f^{IJK}(\overline{\lambda}^J \lambda^K) - f^{IJK} f^{KLM} B^L C^J C^M \doteq 0 \ , \\
& \nabla_a^2 C^I + i f^{IJK}(\overline{\lambda}^J \gamma_9 \lambda^K) - f^{IJK} f^{KLM} B^J B^L C^M \doteq 0 \ ,
\end{aligned} \tag{A.4}$$

which are nothing but superspace reformulation of our component field equations (2.4).

The supersymmetric self-duality conditions for $N = (1/8, 1)$ supersymmetry can now be formulated, first with the restricted fermionic derivative:

$$\widetilde{\nabla}_{\underline{\alpha}} \equiv (\mathcal{P}\nabla)_{\underline{\alpha}} \equiv \mathcal{P}_{\underline{\alpha}\underline{\beta}} \nabla_{\underline{\beta}} \ , \quad \mathcal{P} \equiv \tfrac{1}{8}(P + \tfrac{1}{2} f) \ , \tag{A.5}$$

corresponding to (2.16). From now on, we use the *undotted* $\alpha, \beta, \cdots = 1, 2, \cdots, 8$ for positive chirality fermionic indices, and *dotted* $\dot\alpha, \dot\beta, \cdots = \dot 1, \dot 2, \cdots, \dot 8$ for negative chirality fermionic coordinates: $\underline\alpha = (\alpha, \dot\alpha)$, $\underline\beta = (\beta, \dot\beta)$, $\cdots$.

The supersymmetric self-duality conditions in components (2.15), and (2.19) are now given by the three superfield constraints:

$$\lambda_\alpha{}^I \overset{*}{=} 0 ~, \tag{A.6a}$$

$$\mathcal{P}_{\underline{\alpha}\underline{\beta}} \lambda_{\underline\beta}{}^I \overset{*}{=} 0 ~, \tag{A.6b}$$

$$N_{ab}{}^{cd} F_{cd}{}^I \overset{*}{=} 0 \quad \Longleftrightarrow \quad F_{ab}{}^I \overset{*}{=} \tfrac{1}{2} f_{ab}{}^{cd} F_{cd}{}^I ~, \tag{A.6c}$$

$$\nabla_a(B^I + C^I) \overset{*}{=} 0 \quad \Longleftrightarrow \quad B^I \overset{*}{=} -C^I ~. \tag{A.6d}$$

The consistency among the constraints (A.5), as in component language in section 2, can be easily reconfirmed in superspace, based on the restricted fermionic derivative (A.5). As a typical example, we give the case of $\nabla_\alpha \lambda_\beta \overset{?}{=} 0$:

$$\begin{aligned} 0 \overset{?}{=} \nabla_\alpha \lambda_\beta{}^I &\overset{*}{=} -\tfrac{1}{4}(\mathcal{P}\gamma^{cd})_{\alpha\beta} F_{cd}{}^I - \tfrac{1}{2} f^{IJK} \delta_{\alpha\beta} B^J C^K \\ &\overset{*}{=} -\tfrac{1}{4}(\mathcal{P}\gamma^{cd}_{(+)})_{\alpha\beta} F_{cd}^{(+)I} \equiv 0 \quad (Q.E.D.) ~. \end{aligned} \tag{A.7}$$

The last side vanishes, due to the identity

$$\gamma_{ab}^{(+)} \mathcal{P} \equiv \mathcal{P} \gamma_{ab}^{(+)} \equiv 0 ~, \tag{A.8}$$

which is equivalent to (2.16) and/or (2.17). This identity is confirmed by the use of (2.10) and (2.14). Needless to say, we have to also deal with $\overline\nabla_{\dot\alpha} \lambda_\beta \overset{?}{=} 0$ separately, which turns out to be straightforward under (A.6d). Other consistency confirmation among (A.6) can be also easily performed.

At this point, we reach the stage that such confirmations in superspace looks just parallel to the component language. For this reason, we skip other details such confirmations in this paper.

As a final point, we mention that the closure of supersymmetries is guaranteed to close into translation, even for the self-dual system. This can be easily seen, by recognizing that the supertorsion component $T_{\underline\alpha \underline\beta}{}^c$ is now restricted to

$$T_{\alpha\dot\beta}{}^c = +i(\mathcal{P}\gamma^c)_{\alpha\dot\beta} ~, \quad T_{\dot\alpha\beta}{}^c = +i(\gamma^c \mathcal{P})_{\dot\alpha\beta} ~, \tag{A.9}$$

which is definitely non-vanishing, corresponding to the parameter of translation in (2.23).

We re-emphasize, however, that the confirmation of consistency in superspace provides a nontrivial cross examination of the validity of our total system, such as the commutator algebra, the consistency among supersymmetric self-duality conditions (A.5), together with the reduced Lorentz covariance $Spin(7)$, all reformulated in superspace.